# Characterization of superconducting nanowire single-photon detector with artificial constrictions

Ling Zhang[1, 2](张玲), Lixing You[1,a](尤立星), Dengkuan Liu[1,2](刘登宽), Weijun Zhang[1](张伟君), Lu Zhang[1](张露), Xiaoyu Liu[1](刘晓宇), Junjie Wu[1, 2](巫君杰), Yuhao He[1, 2](何宇昊), Chaolin Lv[1, 2](吕超林), Zhen Wang[1, b](王镇), and Xiaoming Xie[1](谢晓明)

[1]*State Key Laboratory of Functional Materials for Informatics, Shanghai Institute of Microsystem and Information Technology, Chinese Academy of Sciences, 865 Changning Road, Shanghai 200050, China.*

[2]*Graduate University of the Chinese Academy of Sciences, Beijing 100049, P. R. China*

Statistical studies on the performance of different superconducting nanowire single-photon detectors (SNSPDs) on one chip suggested that random constrictions existed in the nanowire that were barely registered by scanning electron microscopy. With the aid of advanced e-beam lithography, artificial geometric constrictions were fabricated on SNSPDs as well as single nanowires. In this way, we studied the influence of artificial constrictions on SNSPDs in a straight forward manner. By introducing artificial constrictions with different wire widths in single nanowires, we concluded that the dark counts of SNSPDs originate from a single constriction. Further introducing artificial constrictions in SNSPDs, we studied the relationship between detection efficiency and kinetic inductance and the bias current, confirming the hypothesis that constrictions exist in SNSPDs.

**I. INTRODUCTION**

Superconducting nanowire single-photon detectors (SNSPDs or SSPDs)[1] have been recognized as a promising technology for ultra-weak optical signal detection. Owing to their excellent broadband sensitivity performance, high detection efficiency (DE), low timing jitter, high count rate, and low dark count rate (DCR), SNSPDs have been applied in many fields, such as free-space optical communications[2], quantum key distribution (QKD)[3, 4], laser ranging[5], and quantum computation [6].

To obtain single-photon detection ability at near-infrared wavelengths, the nanowire linewidth of SNSPDs has to be limited to around 100 nm or even less. Furthermore, many applications require large active areas and multi-arrays of SNSPDs. The uniformity and consistency of large length/breadth ratio nanowires is crucial for SNSPDs. There are two factors that determine the quality of the nanowire. One is the quality of the ultrathin film: both nanoscale defects inside and thickness variations may suppress the quality. The second is geometric constrictions introduced during the fabrication process. In fact, the effects of these two factors are equivalent to each other for SNSPDs. There are several statistical and experimental reports discussing device performance variance[7, 8], which was explained by the possible presence of constrictions in the nanowire, even though they were barely registered.

---

[a] E-mail: lxyou@mail.sim.ac.cn ;
[b] E-mail: zwang@mail.sim.ac.cn .

Owing to advanced nanofabrication technology, it is possible to fabricate a nanowire with artificial constrictions. In this way, we were able to investigate the influence of constrictions on SNSPDs in a straight forward manner. In this study, we designed and fabricated several single nanowires and SNSPDs with different sizes and different numbers of geometric constrictions. The characteristics of the nanowires and SNSPDs are discussed in detail, including the current–voltage curve (I–V), DCR, DE, and kinetic inductance ($L_k$).

## II. FABRICATION AND SAMPLE DESCRIPTION

To study the effect of geometric constrictions, three types of nanowire structures were defined for the study. We defined the width of constriction as the narrowest part in our nanowire. The first type was a single nanowire (120 nm wide and 20 μm long) with a single geometric constriction of five different wire widths (120, 100, 90, 80, and 70 nm). The second was a single nanowire (120 nm wide and 20 μm long) with the same constricted wire size (90 nm wide) but different numbers of constrictions (0, 1, 2, 4, 8). The third type was a traditional SNSPD with a meander structure (15 μm ×15 μm), including one single geometric constriction in the middle of the meandered nanowire. The wire widths of the constrictions were designed to be 100, 90, 80, and 70 nm for four different SNSPDs. The samples were fabricated from a 6.5-nm-thick NbN film deposited on double-sided thermally oxidized silicon substrates. The ultrathin film showed a typical critical temperature, Tc, of ~7 K and a resistivity $\rho_{20k}$ of ~300 μΩ·cm. All nanowire structures were defined by e-beam lithography (EBL) using JEOL JBX-6300FS with e-beam resist PMMA950-A2 and patterned onto NbN by reactive ion etching (RIE) using $CF_4$ and Ar.

Figure 1(a)–(d) shows the scanning electron microscope (SEM) images of a single nanowire sample. To avoid the possible latching effect due to its small inductance, two extra 200-nm-wide meandered nanowires were fabricated in series with the single nanowire (shown as the top and the bottom meanders in Figure 1(a)) to provide extra inductance. There were 8 neighboring nanowires parallel to the single nanowire (shown in Figure 1(b)) to ensure the uniformity of the nanowire linewidth in the EBL process. Figure 1(c) shows four individual artificial geometric constrictions at 1 μm intervals and constriction sizes were defined by EBL. The zoom-in of one artificial constriction is shown in Figure 1(d). The nanowires had an average width of 117 ±2 nm, and the constrictions had a width of 87 ±2 nm and a length of around 100 nm, which was consistent with the design values (90 nm wide and 100 nm long). The width of the nanowire is the average width of 5 different locations on the nanowire measured by SEM, while the width of the constriction was averaged from the measurement result of 5 constrictions with the same design on one chip.

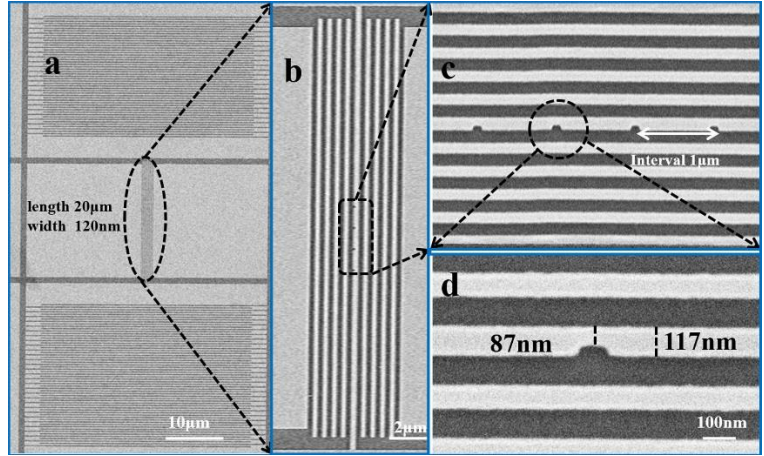

FIG. 1. SEM images of a single nanowire sample. (a) Full SEM image of a single nanowire with two long, wide meandering nanowires in series and 8 neighboring nanowires in parallel. (b) Zoom-in of the single nanowire with 8 neighboring nanowires in parallel. (c) Zoom-in of the nanowire, including four artificial constrictions at 1 μm intervals. (d) Zoom-in of one artificial geometric constriction. The dimension of the constriction and the nanowire is consistent with the design.

## III. RESULTS AND DISCUSSION

All the measurements were carried out in a Gifford–McMahon cryocooler at a working temperature of 2.30 K ±20 mK. The low temperature fluctuation is realized by using a block of stainless steel to damp the original temperature fluctuation (± 140 mK), which ensures the stable operation of SNSPD. To study the intrinsic DCR of the SNSPDs, all samples were packaged inside an oxygen-free copper box fixed to the cold head at the temperature of 2.30 K ±20 mK to avoid any possible thermal radiation or stray light induced dark counts. When we measured the DE of the SNSPDs, single-mode lensed fibers [9] were aligned directly to the SNSPDs to ensure good optical coupling. The electronic response pulse of SNSPD was amplified by a room temperature low noise amplifier (LNA-650 from RF-Bay inc), and then fed into a photon counter (SR400 from Stanford Research Systems) for counting.

### A. SINGLE NANOWIRE WITH ARTIFICIAL CONSTRICTIONS

To study the relationship between the intrinsic DCR and constrictions, a single nanowire instead of a meandered nanowire was chosen to minimize the possibility of random constrictions caused by either poor film quality or process error. Figure 2 shows the I–V curves of five nanowires with artificial constrictions of different widths (120, 100, 90, 80, and 70 nm from top to bottom). All of the I–V curves showed a consistent behavior, except for variations in the switching current ($I_{sw}$). The inset of Figure 2 shows the width ($W_c$) dependence of $I_{sw}$. The linear equation $I_{sw} = J_{sw} \cdot d \cdot W_c$ with $J_{sw}$ = 3.8 ±0.4 MA/cm$^2$ fits the experimental values well, which indicates a feasible control of the fabrication process. However, this switching current density $J_{sw}$ may not represent the intrinsic critical current density. A series of articles have proven that current-crowding effect appears in the bend or the turn in a superconducting nanowire which happens to exist in artificial constrictions[10-13]. As a result, $J_{sw}$ should be smaller than the intrinsic critical current density of the film.

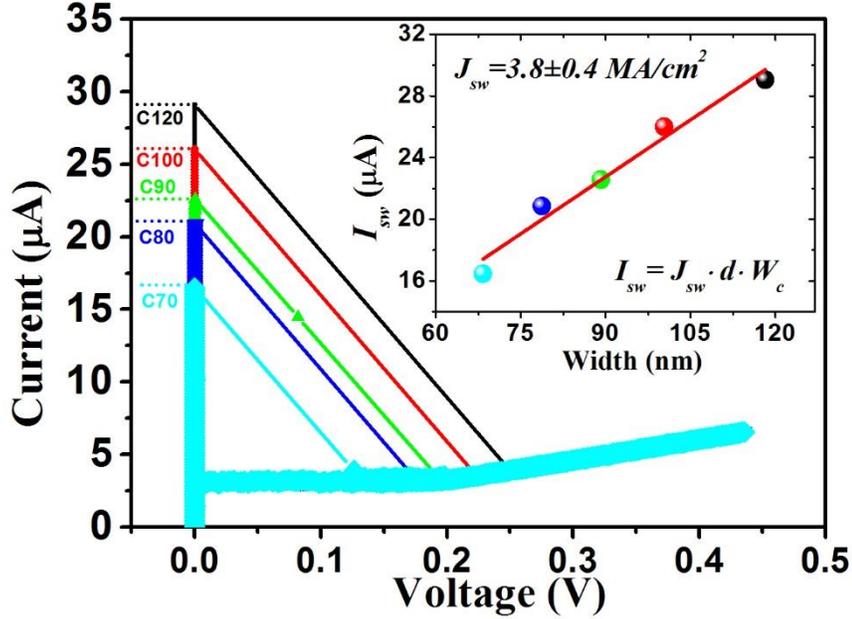

FIG. 2. I–V curves of five single nanowires (120 nm wide and 20 μm long) with constrictions of different widths. I–V curves from top to bottom correspond to the width of the constrictions of 120 (i.e., no constriction), 100, 90, 80, and 70 nm, respectively. The inset shows the relationship between $I_{sw}$ and $W_c$. The red line indicates the linear fitting equation $I_{sw} = J_{sw} \cdot d \cdot W_c$, where $d=6.5$ nm is the thickness of the nanowire.

The inset of Figure 3 shows the current dependence of the DCR of the nanowires described above. As the samples were thermally and optically shielded, we did not see any background dark counts originating from either blackbody radiation or stray light. As a result, the DCR we measured represented the intrinsic DCR of the samples. We noticed that all the relationships had the same linear behavior, but with different current intercepts. We normalized the bias current $I_b$ by its switching current and all the $DCR - I_b/I_{sw}$ curves of the nanowire with different constriction sizes showed the same slope (Figure 3).

For a single nanowire with an artificial constriction, the contribution of the dark counts may be divided into two parts. One part is the dark counts from the artificial constriction; the other part is the sum of the dark counts from all other areas along the nanowire. As there were no artificial constrictions in nanowire C120, the DCR of C120 (square dots in the inset of Figure 3) may represent the second part of the contribution, because the length of the constriction is negligible compared with the length of the nanowire. If we examine the DCR of C120 with bias currents between 24 and 26 μA (typical current values for C100; diamond dots in the inset of Figure 3), the value is far below $10^{-3}$ Hz. This indicates that the second part of the DCR, i.e. DCR from all the other area along the nanowire without constriction, is negligible for C100. In the other words, all the dark counts in C100 originated from the artificial 100-nm-wide constriction. Likewise, all the dark counts in C90, C80, and C70 were contributed from the single artificial constriction of the corresponding size. Indeed, even though there was no artificial constrictions in C120, either the terminal of the nanowire (the 90-degree turns) or some possible random constriction (in width or thickness) would also cause some current-crowding effect[14], which functioned similar as the existence of a geometrical constriction. This suggests that the dark counts in SNSPDs originate from constrictions in the nanowire.

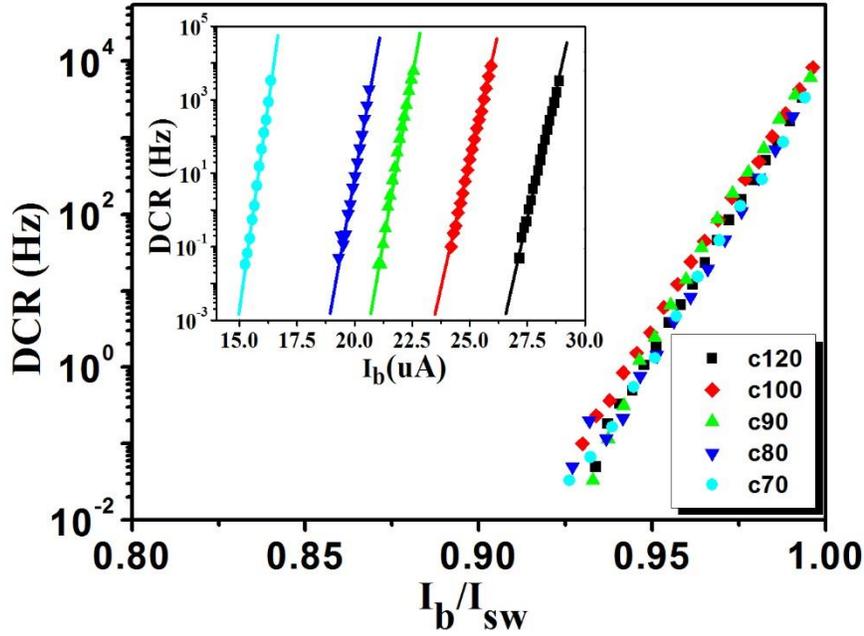

FIG. 3. DCR as a function of the bias current normalized by the switching current. The inset shows the DCR of the SNSPDs with different constriction widths as a function of the bias current. The straight lines in the inset are provided as a guide to the eye.

Based on the above results, we speculate that the DCR should increase with the number of the constrictions with the same size, as long as there is no correlation between them. In fact, assuming a maximum dark count rate of $10^4$ Hz and a dark count pulse duration of 100 ns, the sum duration of the dark counts per second is 1 ms for a single constriction. As a result, we may consider all the dark counts as individual events in the time domain for a nanowire with a few constrictions. To verify the idea, nanowires with different numbers of constrictions with the same design width were fabricated. Figure 1(c) shows a 120-nm-wide nanowire including four constrictions with a width of ~90 nm. The interval between the constrictions was set to be 1 μm to avoid possible geometrical correlation, as the typical hotspot in SNSPDs has a size of around 100 nm[15]. We compared the current dependence of the DCR for the nanowires with 0, 1, 2, 4, 8 constrictions. We did not see any relationship between the DCR and the number of constrictions. In fact, the results were similar to the data in Figure 3.

By examining the geometric precision of the nanowire and constrictions, we noticed that there was a ~3% distribution in the linewidth, which would cause an uncertainty of 3% in $I_{sw}$ of the nanowires. Besides, the thickness fluctuation of the constrictions may also contribute to the uncertainty. For example, the typical roughness of 0.2 nm (measured by atomic force microscope) for 6.5 nm-thick NbN film may result in an uncertainty of 6% in $I_{sw}$. For $I_b/I_{sw}$ in the range of 0.9 to 1.0, uncertainty of 3% in $I_{sw}$ corresponds to the uncertainty of 3% in $I_b/I_{sw}$. From the data in Figure 3, the changes of 3% in $I_b/I_{sw}$ results in two orders of magnitude difference in the DCR. This explains why we did not observe a larger DCR in the multi-constriction experiments. To successfully demonstrate this experiment, we need to control the variance of the nanowire linewidth and the film roughness to be <1%, which is still a challenge in state-of-art fabrication technology.

It is noted here that, though we suggested that the dark counts in SNSPD originate from the constriction in the nanowire, the physical mechanism of the dark counts is still not well determined. There are three possible mechanisms related to the dark counts in SNSPD, which include vortex-antivortex pairs, vortex hopping and quantum tunneling[16-20]. By studying the temperature and line-width dependences of the dark count rate, a recent article indicated that neither sources of dissipation involving vortex-antivortex pairs, vortex hopping and quantum tunneling can be neglected for the contribution to the dark count rate[21].

## B. SNSPDS WITH ARTIFICIAL CONSTRICTIONS

By studying the characteristics of single nanowires with constrictions of different sizes, we deduced that the dark counts in SNSPDs originate from the constrictions. However, it is difficult to study DE of a single nanowire due to the low optical absorptance. Instead, we fabricated constrictions with different sizes in classical meander-structured SNSPDs. In this way, we were able to study the effect of artificial constrictions on the DE.

Three different SNSPDs were fabricated, in which a constriction with different widths (90, 80, and 70 nm) was designed on the center nanowire. The designed nanowire width of the SNSPDs was 100 nm. One normal SNSPD without any artificial constriction was also fabricated for comparison. Practically, the widths of the nanowire and the constrictions were slightly smaller than the designed value, which were 98, 85, 73, and 62 nm (with a standard deviation of ±2 nm concluded by 3 random measurements of the same constriction), as observed by SEM. The practical width ratios ($R_w$) of the constrictions to the nanowire were 1.00, 0.86, 0.75, and 0.63. In Figure 4, C98, C85, C73, and C62 represent the devices described in Figure 4. The ratio of the bias currents for a device with a constriction to a device without a constriction at the same DE (0.1%) can be defined as parameter $C$[7]. The relationship between the DE and the normalized biased current $I_b/I_{sw} \cdot C$ is shown in Figure 4. All four curves matched each other well at high current values. However, at low current values, a deviation appeared for the two SNSPDs with smaller constrictions, which can be explained by the contribution of the DE from the constriction area in the total DE. The results are consistent with previous results[7, 8]. Interestingly, the $C$ values for the curves are same as the calculated values of $R_w$, which is the defined geometric parameter for the constriction. This result directly suggests that the variations in DE are indeed caused by the constrictions.

Similarly, we examined the current dependence of the kinetic inductance $L_k$ for the above SNSPDs. $L_k$ was obtained from the measurement of the phase of the reflection coefficient S11 as a function of frequency[22]. Figure 5 shows the current dependence of the $L_k$ ratio, which is expressed as $L_k(I)/L_k(0)$. For a typical SNSPD without a constriction, the $L_k$ ratio should increase as the bias current increases, finally reaching around 1.2 at the switching current[7]. In our measurements, the value of the ratio reached 1.23 for the no-constriction device C98 (black dotted line in Figure 5). However, for the constriction widths of 85, 73, and 62 nm, the $L_k$ ratio was 1.17, 1.12, and 1.07, respectively. The increase in the $L_k$ ratio is suppressed by the constrictions, which prevents $I_b$ from approaching the $I_{sw}$ of C100.

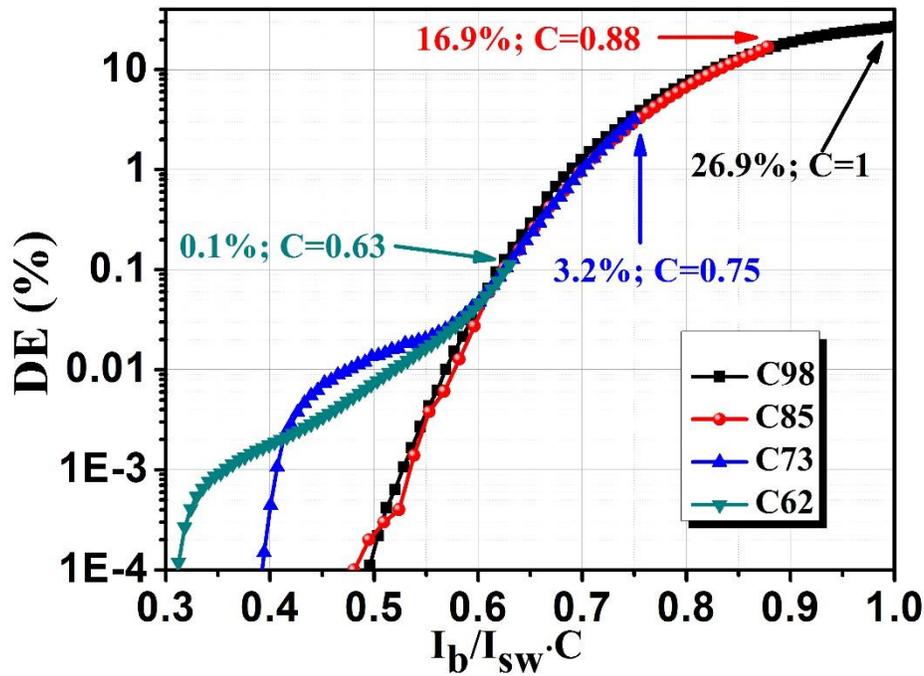

FIG. 4. DE of devices with different constriction sizes versus the normalized biased current. The values of $C$ and the maximal DE are indicated. The lines are provided as a guide to the eye.

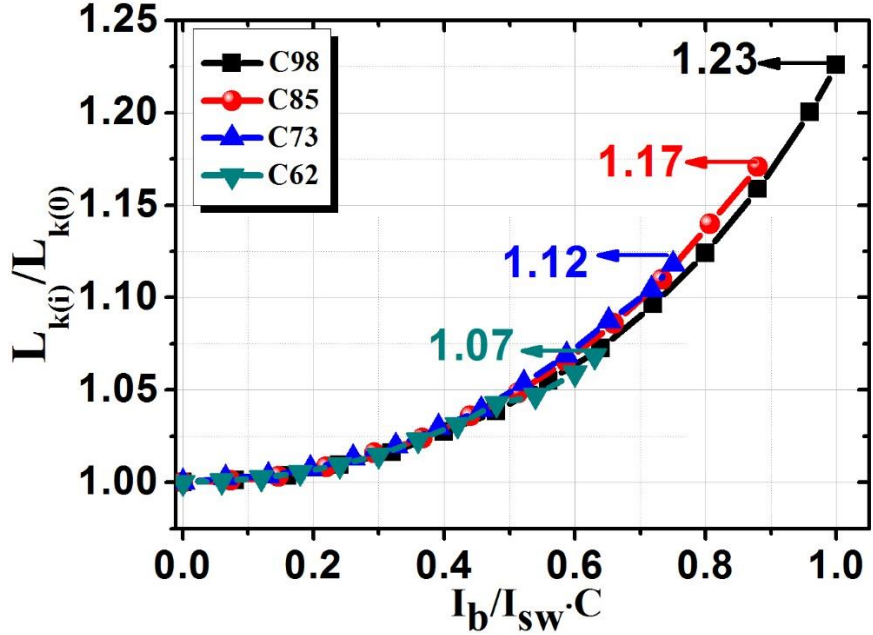

FIG. 5. Current dependence of the normalized kinetic inductance. $L_k(0)$ represents the inductance of SNSPDs at 2.3 K without the current bias. The lines in the inset are provided as a guide to the eye.

## IV. CONCLUSION

Single nanowires and SNSPDs with artificial constrictions were fabricated. We extensively studied the influence of the constrictions on SNSPDs. From the DCR characteristics of the single nanowires with constrictions of different widths, we suggested that the dark counts in SNSPDs originate in the constriction. From SNSPDs with constrictions of different widths, we found consistent normalized curves for both DE and $L_k$, which can be explained by the presence of constrictions. Our results of several samples are consistent with the statistical results of over 100 samples in previous reports[7, 23], which finds evidence for the previous hypothesis that constrictions exist in practical SNSPDs.

## ACKNOWLEDGEMENTS

The authors would like to acknowledge Taro Yamashita for fruitful discussions on the origin of dark counts. This work was supported by the National Natural Science Foundation of China (91121022), Strategic Priority Research Program (B) of the Chinese Academy of Sciences (XDB04010200 and XDB04020100), the National Basic Research Program of China (2011CBA00202), and the National High-Tech Research and Development Program of China (2011AA010802).